\documentclass[12pt]{article}
\pdfoutput=1
\usepackage[left=1.25in,top=1in]{geometry}
\usepackage{graphicx,url}
\usepackage{tabularx}
\usepackage[small,bf]{caption}

\makeatletter
\renewcommand{\paragraph}{%
  \@startsection{paragraph}{4}%
  {\z@}{1.25ex \@plus 1ex \@minus .2ex}{-1em}%
  {\normalfont\normalsize\bfseries}%
}
\makeatother

\newtheorem{theorem}{Theorem}

\newcommand{\Oh}[1]
  {\ensuremath{\mathcal{O}\!\left( {#1} \right)}}
\newcommand{\forward}
  {\ensuremath{\mathsf{forward}}}
\newcommand{\backward}
  {\ensuremath{\mathsf{backward}}}
\newcommand{\lastchar}
  {\ensuremath{\mathsf{lastchar}}}
\newcommand{\shorter}
  {\ensuremath{\mathsf{shorter}}}
\newcommand{\longer}
  {\ensuremath{\mathsf{longer}}}
\newcommand{\maxlen}
  {\ensuremath{\mathsf{maxlen}}}
\newcommand{\rank}
  {\ensuremath{\mathsf{rank}}}
\newcommand{\select}
  {\ensuremath{\mathsf{select}}}

\newcommand{\nodelabel}
  {\ensuremath{\mathsf{label}}}

\begin{document}

\title{\vspace{-4ex}Variable-Order de Bruijn Graphs}
\author{\normalsize Christina Boucher$^1$, Alex Bowe$^2$, Travis Gagie$^3$,\\
\normalsize Simon J.\ Puglisi$^3$ and Kunihiko Sadakane$^4$\\[1ex]
\footnotesize $^1$ Department of Computer Science, Colorado State University, Fort Collins, CO\\[-0.8ex]
\footnotesize $^2$ National Institute of Informatics, Japan\\[-0.8ex]
\footnotesize $^3$ Helsinki Institute for Information Technology,\\[-0.8ex]
\footnotesize Department of Computer Science, University of Helsinki, Finland\\[-0.8ex]
\footnotesize $^4$ School of Information Science and Technology, University of Tokyo, Japan}
\date{}
\maketitle

\begin{abstract}
The de Bruijn graph $G_K$ of a set of strings $S$ is a key data structure in 
genome assembly that represents overlaps between all the $K$-length substrings 
of $S$. Construction and navigation of the graph is a space and time bottleneck 
in practice and the main hurdle for assembling large, eukaryote genomes. 
This problem is compounded by the fact that state-of-the-art assemblers do not build the de Bruijn graph for a single order (value of $K$) but for multiple values of $K$.
More precisely, they build or update $d$ de Bruijn graphs, each with a specific order, i.e., $G_{K_1}, G_{K_2}, \ldots, G_{K_d}$.  
Although, this paradigm increases the quality of the assembly produced, it increases the run time by roughly a factor of $d$.
In this paper, we show how to augment a succinct de Bruijn graph 
representation by Bowe et al. (Proc. WABI, 2012) 
to support new operations that let us change order on the fly, effectively
representing all de Bruijn graphs of order up to some maximum $K$ in
a single data structure. 
Our experiments show our variable-order de Bruijn graph only modestly increases
space usage, construction time, and navigation time compared to a single order graph.
\end{abstract}

\section{Introduction}
\label{sec:introduction}

Accurate assembly of genomes is a fundamental problem in bioinformatics and is vital to several ambitious scientific projects, including the 10,000 vertebrate genomes (Genome 10K)~\cite{Haussler09},  \emph{Arabidopsis} variations (1001 genomes)~\cite{Ossowski08}, human variations (1000 genomes)~\cite{Abecasis12}, and the Human Microbiome~\cite{hmp} projects. The genome assembly process builds long contiguous DNA sequences, called {\em contigs}, from shorter DNA fragments, called {\em reads}, typically 100-150 (DNA) symbols in length. 

In Eulerian sequence assembly \cite{IW95,PTW}, a {\em de Bruijn graph} is constructed with a vertex $v$ for every $K$-mer (substring of length $K$) present in an input set of reads, and an edge $(v, v')$ for every observed $(K + 1)$-mer in the reads with $K$-mer prefix $v$ and $K$-mer suffix $v'$. Each contig corresponds to a non-branching path through this graph. Most state-of-the-art assemblers use this paradigm~\cite{bankevich2012spades,peng2010idba,Li:2010,Simpson:2009,Butler:2008}, 
and follow the same general outline: extract $(K + 1)$-mers from the reads; construct the de Bruijn graph on the set of $(K +1)$-mers; simplify the graph; 
and construct the contigs (simple paths in the graph). The value of $K$ can be, and is often required to be, specified by the user.

Determining an appropriate value of $K$ is important and has a direct impact on assembly quality. Stated very briefly, when $K$ is too small the resulting graph is complicated by spurious edges and nodes, and when $K$ is too large the graph becomes too sparse and possibly disconnected.
Repetitive regions in the underlying genome are especially problematic since they give rise to spurious edges and nodes in the de Bruijn graph \cite{alkan2011limitations} and are very sensitive to the choice of $K$.

In an attempt to circumvent the need to choose a single, ideal value of $K$, SPAdes~\cite{bankevich2012spades} and IDBA~\cite{peng2010idba} use a number
of different $K$ values.
IDBA~\cite{peng2010idba} builds a number of de Bruijn graphs for each a fixed set of $K$ values. 
At a given iteration of the algorithm, the de Bruin graph for the current value of $K$ is built from the reads and 
the contigs for that graph are constructed, then all the reads that align to at least one of those contigs are removed from the current 
set of reads. In the next iteration the graph is built by converting every edge from the previous graph to a vertex while treating contigs 
as edges. SPAdes \cite{bankevich2012spades} uses a similar approach but uses all the reads at each iteration.  

\paragraph{Our Contribution.}
SPAdes~\cite{bankevich2012spades} and IDBA~\cite{peng2010idba} represent the state-of-the-art for 
genome assemblers, producing assemblies of greatly improved quality compared to previous approaches. However, 
their need to construct several de Bruijn graphs of different orders over the assembly process makes them 
extremely slow on large genomes. In this paper we address this problem by describing a succinct data structure
that, for a given $K$, efficiently represents {\em all} the de Bruijn graphs for $k \le K$
and allows navigation within and between each graph. We have implemented our data structure and show that in
practice it requires around twice the space of a graph for a single $K$, 
and incurs a modest slow down in construction time and on navigation operations.

\paragraph{Related Work.} There are several succinct data structures for the de Bruijn graph of a single order (i.e.~value of $K$).  One of the first approaches was introduced by Simpson et al.~\cite{Simpson:2009} as part of the development of the ABySS assembler.  Their method stores the graph as a distributed hash table and thus requires 336G to store the graph corresponding to a set of reads from a human genome (HapMap: NA18507).  In 2011, Conway and Bromage~\cite{conway} reduced space requirements 
by using a sparse bitvector (by Okanohara and Sadakane \cite{bitvector}) to represent the $(K + 1)$-mers (the edges), and used rank and select operations (to be described shortly) to traverse it. As a result, their representation took 32 GB for the same data set.  Minia, by Chikhi and Rizk \cite{wabi}, uses a bloom filter
to store edges.
They traverse the graph by generating all possible outgoing edges at each node and testing their membership in the bloom filter. Using this approach, the graph was reduced to 5.7 GB on the same dataset.  Contemporaneously, Bowe, Onodera, Sadakane and Shibuya~\cite{bowe} developed a different succinct data structure based on the Burrows-Wheeler transform~\cite{bw1994} that requires 2.5 GB.  Their representation, which henceforth we refer to as BOSS from the authors' initials, is a starting point for our methods and we will discuss it in detail below.

The data structure of Bowe et al.~\cite{bowe}
is combined with ideas from IDBA-UD~\cite{idbaud} in a metagenomics assembler called MEGAHIT~\cite{megahit}. 
In practice MEGAHIT requires more memory than competing methods 
but produces significantly better assemblies.

Lastly, Lin and Pevzner~\cite{mdbg} recently introduced the {\em manifold de Bruijn graph}, which 
associates arbitrary substrings with nodes (the substrings are fixed during preprocessing), 
rather than $K$-mers. Lin and Pevzner's structure is mainly of theoretical interest since it has not yet been implemented.


\paragraph{Roadmap.} Section \ref{sec:preliminaries} sets notation, and formally lays down the problem and auxiliary data structures we use.   Section~\ref{sec:BOSS} gives details of the BOSS representation~\cite{bowe}. Section~\ref{sec:changing} then describes our variable-order de Bruijn graph structure. In Section~\ref{sec:experiments} we report on experiments comparing the practical performance of our data structure to that of a single-order de Bruijn graph. Section~\ref{sec:conclusion} offers directions for future work.

\section{Preliminaries}
\label{sec:preliminaries}

\paragraph{De Bruijn Graphs.}
\label{sec:dbg}

Given an alphabet $\Sigma$ of $\sigma$ symbols and a set of strings 
$\lbrace S_1, S_2, \ldots, S_t \rbrace$, $S_i \in \Sigma^{+}$, the 
{\em de Bruijn graph} of order $K$, denoted $G^S_K$, or just $G_K$, 
when the context is clear, is a directed, labelled graph defined as 
follows.

Let $M_{K}$ be the set of distinct $K$-mers (strings of length $K$) 
that occur as substrings of some $S_i$. $M_{K+1}$ is defined similarly. 
$G_K$ has exactly $|M_{K}|$ nodes and with each node $u$ we associate 
a distinct $K$-mer from $M_{K}$, denoted $\nodelabel(u)$. Edges are 
defined by $M_{K+1}$: for each string $T \in M_{K+1}$ 
there is a directed edge, labelled with symbol $T[K+1]$, from node 
$u$ to node $v$, where $\nodelabel(u) = T[1,K]$ and $\nodelabel(v) = T[2,K+1]$. 

\paragraph{Rank and Select.}
\label{sec:rank}
Two basic operations
used in almost every succinct and compressed data structure are {\em rank} and
{\em select}. Given a sequence (string) $S[1,n]$ over an alphabet $\Sigma =
\{1,\ldots,\sigma\}$, a character $c \in \Sigma $, and integers
$i$,$j$, $\rank_c(S,i)$ is the number of times that $c$ appears in
$S[1,i]$, and $\select_c(S,j)$ is the position of the $j$-th
occurrence of $c$ in $S$.
For a binary string $B[1,n]$, the classic solution for rank and select~\cite{Mun96} 
is built upon the input sequence, requiring $o(n)$ additional bits.
Generally, $\rank_1$ and $\select_1$ are considered the default
rank and select queries.
More advanced solutions (e.g.~\cite{bitvector}) achieve zero-order 
compression of $B$,
representing it in just $nH_0(B) + o(n)$ bits of space, and supporting $\rank$ and
$\select$ operations in constant time. 

\paragraph{Wavelet Trees.}
\label{sec:WVT}
To support rank and select on larger alphabet strings, the wavelet tree~\cite{ggv2003,n2013} is a 
commonly used data structure that occupies $n\log\sigma + o(n\log\sigma)$
bits of space and supports $\rank$ and $\select$ queries in $\Oh{\log\sigma}$ time.
Wavelet trees also support a variety of more complex queries on the underlying string (see, e.g.~\cite{gnp2012}),
in $\Oh{\log\sigma}$ time, and we will make use of some of this functionality in Section~\ref{sec:implementations}.

\section{BOSS representation}
\label{sec:BOSS}

Conceptually, to build the BOSS representation~\cite{bowe} of a $K$th-order de Bruijn graph from a set of \((K + 1)\)-tuples, we first add enough dummy \((K + 1)\)-tuples starting with \$s so that if \(\alpha a\) is in the set, then some tuple ends with $\alpha$.  We also add enough dummy \((K + 1)\)-tuples ending with \$ that if \(b \alpha\) is in the set, with $\alpha$ containing no \$ symbols, then some tuple starts with $\alpha$.  We then sort the set of \((K + 1)\)-tuples into the right-to-left lexicographic order of their first $K$ characters (with ties broken by the last character) to obtain a matrix.  If the $i$th through $j$th \((K + 1)\)-tuples start with $\alpha$, then we say node \([i, j]\) in the graph has label $\alpha$, with \(j - i + 1\) outgoing edges labelled with the last characters of the $i$th through $j$th \((K + 1)\)-tuples.  If there are $n$ nodes in the graph, then there are at most \(\sigma n\) rows in the matrix, i.e., \((K + 1)\)-tuples.

\begin{figure}
\begin{center}
\begin{tabular}{c@{\hspace{10ex}}c}
\begin{tabular}{r@{\hspace{1ex}}@{\hspace{1ex}}@{\hspace{1ex}}l@{\hspace{1ex}}c}
1) & \,\$\,\$\,\$\, & T\\
2) & CGA & C\\
3) & \,\$\,TA & C\\
4) & GAC & G\\
5) & GAC & T\\
6) & TAC & G\\
7) & GTC & G\\
8) & ACG & A\\
9) & ACG & T\\
10) & TCG & A\\
11) & \,\$\,\$\,T & A\\
12) & ACT & \$\\
13) & CGT & C
\end{tabular} &
\raisebox{-10ex}
{\includegraphics*[trim = 0cm 0cm 9cm 24cm, width=50ex]{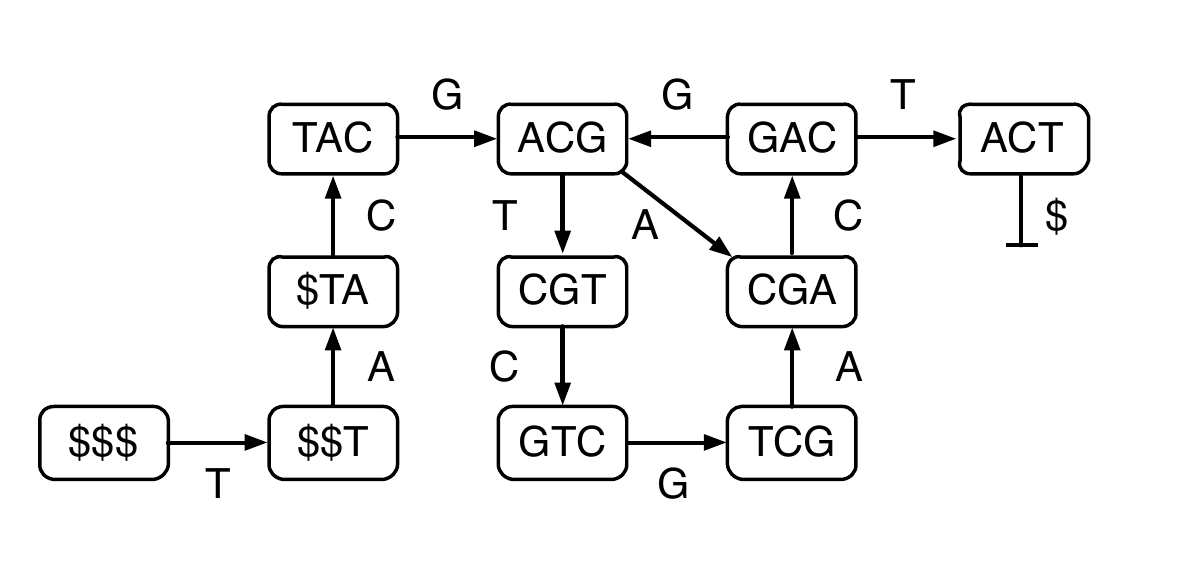}}
\end{tabular}
\caption{The BOSS matrix (left) and de Bruijn graph (right) for the quadruples CGAC, GACG, GACT, TACG, GTCG, ACGA, ACGT, TCGA, CGTC.}
\label{fig:matrix}
\end{center}
\end{figure}

For example, if \(K = 3\) and the matrix is the one from Bowe et al.'s paper, shown in the left of Figure~\ref{fig:matrix}, then the \(n = 11\) nodes are
\[[1, 1], [2, 2], [3, 3], [4, 5], [6, 6], [7, 7], [8, 9], [10, 10], [11, 11], [12, 12], [13, 13]\]
with labels
\[\mathrm{\$\$\$, CGA, \$TA, GAC, TAC, GTC, ACG, TCG, \$\$T, ACT, CGT}\,,\]
respectively.  The 3rd-order de Bruijn graph itself is shown in the right of the figure.

Bowe et al.\ described a number of queries on the graph, all of which can be implemented in terms of the following three with at most an $\Oh{\sigma}$-factor slowdown:
\begin{itemize}
\item $\forward(v, a)$ returns the node $w$ reached from $v$ by an edge labelled $a$, or NULL if there is no such node;
\item $\backward(v)$ lists the nodes $u$ with an edge from $u$ to $v$;
\item $\lastchar(v)$ returns the last character of $v$'s label.
\end{itemize}
In our example, \(\forward([8, 9], \mathrm{A}) = [2, 2]\), \(\backward([2, 2]) = [8, 9], [10, 10]\) and \(\lastchar \allowbreak ([8, 9]) = \mathrm{G}\).  Since $\backward$ always returns at least one node, we can recover any non-dummy node's entire label by $K$ calls to $\lastchar$ interleaved with \(K - 1\) calls to $\backward$.

\section{Changing order}
\label{sec:changing}

If we delete the first column of the matrix in Figure~\ref{fig:matrix}, the result is {\em almost} the BOSS matrix for a 2nd-order de Bruijn graph whose nodes
\[[1, 1], [2, 2], [3, 3], [4, 6], [7, 7], [8, 10], [11, 11], [12, 12], [13, 13]\]
have labels
\[\mathrm{\$\$, GA, TA, AC, TC, CG, \$T, CT, GT}\,,\]
respectively.  Similarly, if we delete the first two columns of the original matrix, the result is almost the BOSS matrix for a 1st-order graph whose nodes
\[[1, 1], [2, 3], [4, 7], [8, 10], [11, 13]\]
have labels
\[\mathrm{\$, A, C, G, T}\,,\]
respectively.  If we delete the first three columns, the result is almost the BOSS graph for the 0th-order graph whose single node \([1, 13]\) has an empty label.  Notice we allow the same node to appear in different graphs, with labels of different lengths.  If readers find this confusing, they can imagine that nodes are triples instead of pairs, with the additional component storing the label's length.

The truncated form of a higher order BOSS differs from the BOSS of a lower order in that
some rows are repeated, which could prevent the BOSS representation from working properly.  Suppose that, instead of trying to apply $\forward$, $\backward$ and $\lastchar$ directly to nodes in the new graphs, we augment the BOSS representation of the original graph to support the following three queries:
\begin{itemize}
\item $\shorter(v, k)$ returns the node whose label is the last $k$ characters of $v$'s label;
\item $\longer(v, k)$ lists nodes whose labels have length \(k \leq K\) and end with $v$'s label;
\item $\maxlen(v, a)$ returns some node in the original graph whose label ends with $v$'s label, and that has an outgoing edge labelled $a$, or NULL otherwise. 
\end{itemize}
If we want a node in the original graph whose label ends with $v$'s label but we do not care about its outgoing edges, then we write \(\maxlen(v, *)\).  Notice $\shorter$ and $\longer$ are symmetric, in the sense that if $v$'s label has length $k_v$ and \(x \in \longer(v, k_x)\), then \(\shorter(x, k_v) = v\).  In our example, \(\shorter([4, 5], 2) = [4, 6]\) while \(\longer([4, 6], 3) = [4, 5], [6, 6]\) and \(\maxlen ([4, 6], \mathrm{G})\) could return either \([4, 5]\) or \([6, 6]\), while \(\maxlen([4, 6], \mathrm{T}) = [4, 5]\) and \(\maxlen([4, 6], \mathrm{A}) = \mathrm{NULL}\).

If $v$ is a node in the original graph --- e.g., $v$ is returned by $\maxlen$ --- then we can use the BOSS implementations of $\forward$, $\backward$ and $\lastchar$.  Otherwise, if $v$'s label has length $k_v$ then
\begin{eqnarray*}
\forward(v, a) & = & \shorter(\forward(\maxlen(v, a), a), k_v)\\
\lastchar(v) & = & \lastchar(\maxlen(v, *))\,.
\end{eqnarray*}
Assuming queries can be applied to lists of nodes, we can compute \(\backward(v)\) as 
\[\shorter(\backward(\maxlen(\longer(v, k_v + 1), *)), k_v),\]
removing any duplicates.

To see why we can compute $\backward$ like this, suppose $v$'s label is \(\alpha a\), so \(\longer(v, \allowbreak k_v + 1)\) returns a list of all \(d \leq \sigma\) nodes whose labels have the form \(b \alpha a\).  Applying $\maxlen$ to this list returns a second list of $d$ nodes, with labels \(\beta_1 b_1 \alpha a, \ldots, \beta_d b_d \alpha a\) of length $K$.  Applying $\backward$ to this second list returns yet a third list, of all the at most \(\sigma d\) nodes whose labels have the form \(c \beta_i b_i \alpha\).  We need only one node returned calling $\backward$ on each node in the second list, so we can discard all but at most $d$ nodes in the third list.  Finally, applying $\shorter$ to the third list returns a fourth list, of all $d$ nodes whose labels have the form \(b_i \alpha\), each of which may be repeated at most $\sigma$ times in the list.

\section{Implementing $\shorter$, $\longer$ and $\maxlen$}
\label{sec:implementations}

The BOSS representation includes a wavelet tree over the last column $W$ of the BOSS matrix, and a bitvector $L$ of the same length with 1s marking where nodes' intervals end.  In our example, \(W = \mathrm{TCCGTGGATAA\$C}\) and \(L = 1110111011111\).

Now we can implement \(\maxlen([i, j], a)\) in $\Oh{\log \sigma}$ time: we use $\rank$ and $\select$ on $W$ to find an occurrence \(W [r]\) of $a$ in \(W [i..j]\), if there is one; we then use $\rank$ and $\select$ on $L$ to find the last bit \(L [i' - 1] = 1\) with \(i' \leq r\) and the first bit \(L [j'] = 1\) with \(j' \geq r\), and return \([i', j']\).  (If there is no occurrence of 1 strictly before \(L [r]\), then we set \(i' = 1\).)  We can implement \(\maxlen([i, j], *)\) in $\Oh{1}$ time: instead of using $\rank$ and $\select$ on $W$ to find $r$, we simply choose any $r$ between $i$ and $j$.

In our example, for \(\maxlen([4, 6], \mathrm{G})\) we first find an occurence \(W [r]\) of G in \(W [4..6]\), which could be either \(W [4]\) or \(W [6]\); if we choose \(r = 4\) then the last bit \(L [i' - 1] = 1\) with \(i' \leq r\) is \(L [3]\) and the first bit \(L [j'] = 1\) with \(j' \geq r\) is \(L [5]\), so we return \([i', j'] = [4, 5]\); if we choose \(r = 6\) then the last bit \(L [i' - 1] = 1\) with \(i' \leq r\) is \(L [5]\) and the first bit \(L [j'] = 1\) with \(j' \geq r\) is \(L [6]\), so we return \([i', j'] = [6, 6]\).

To implement $\shorter$ and $\longer$, we store a wavelet tree over the sequence $L^*$ in which \(L^* [i]\) is the length of the longest common suffix of the label of the node in the original graph whose interval includes $i$, and the label of the node whose interval includes \(i + 1\); this takes $\Oh{\log K}$ bits per \((K + 1)\)-tuple in the matrix.  To save space, we can omit $K$s in $L^*$, since they correspond to 0s in $L$ and indicate that $i$ and \(i + 1\) are in the interval of the same node in the original graph; the wavelet tree then takes $\Oh{\log K}$ bits per node in the original graph and $\Oh{n \log K}$ bits in total.  In our example, \(L^* = 0, 1, 0, 3, 2, 1, 0, 3, 2, 0, 1, 1\) (and we can omit the 3s to save space).

For \(\shorter([i, j], k)\), we use the wavelet tree over $L^*$ to find the largest \(i' \leq i\) and the smallest \(j' \geq j\) with \(L^* [i' - 1], L^* [j'] < k\) and return \([i', j']\), which takes $\Oh{\log K}$ time.  For \(\longer([i, j], k)\), we use the wavelet tree to find the set \(B = \{b\,:\,L^* [b] < k\,;\,i - 1 \leq b \leq j\}\) --- which includes \(i - 1\) and $j$ --- and then, for each consecutive pair \((b, b')\) in $B$, we report \([b + 1, b']\); this takes a total of $\Oh{|B| \log K}$ time.  With these implementations, if the time bounds for \(\forward(v, a)\), \(\backward(v)\) and \(\lastchar(v)\) are $\Oh{t_\forward}$, $\Oh{t_\backward}$ and $\Oh{t_\lastchar}$ when $v$ is a node in the original graph, respectively, then they are $\Oh{t_\forward + \log \sigma + \log K}$, $\Oh{\sigma (t_\backward + \log K)}$ and $\Oh{t_\lastchar + 1}$ when $v$ is not a node in the original graph.

In our example, for \(\shorter([4, 5], 2)\) we find the largest \(i' \leq 4\) and the smallest \(j' \geq 5\) with \(L^* [i' - 1], L^* [j'] < 2\) --- which are 4 and 6, respectively --- and return \([4, 6]\).  For \(\longer([4, 6], 3)\) we find the set \(B = \{b\,:\,L^* [b] < 3\,;\,3 \leq b \leq 6\} = \{3, 5, 6\}\) and report \([4, 5]\) and \([6, 6]\).

A smaller but slower approach is not to store $L^*$ explicitly but to support access to any cell \(L^* [i]\) by finding the nodes in the original graph whose intervals include $i$ and \(i + 1\), then using $\backward$ and $\lastchar$ to compute their labels and find the length of their longest common suffix; this takes a total of $\Oh{K (t_\backward + t_\lastchar)}$ time.  To implement $\shorter$ and $\longer$, we store a range-minimum data structure~\cite{fh2011} over $L^*$, which takes \(2 n + o (n)\) bits and returns the position of the minimum value in a specified substring of $L^*$ in $\Oh{1}$ time.

For \(\shorter([i, j], k)\), we use binary search and range-minimum queries to find the largest \(i' \leq i\) and the smallest \(j' \geq j\) with \(L^* [i' - 1], L^* [j'] < k\) and return \([i', j']\), which takes $\Oh{K (t_\backward + t_\lastchar) \log (n \sigma)}$ time.  (With a more complicated use of the range-minimum data structure, which we will describe in the full version of this paper, we use $\Oh{K^2 (t_\backward + t_\lastchar)}$ time.)  For \(\longer([i, j], k)\), we recursively split \([i, j]\) into subintervals with range-minimum queries, at each step using $\backward$ and $\lastchar$ to check that the minimum value found is less than $k$; this takes $\Oh{K (t_\backward + t_\lastchar)}$ time per node returned.  With these implementations, \(\forward(v, a)\), \(\backward(v)\) and \(\lastchar(v)\) take $\Oh{t_\forward + K (t_\backward + t_\lastchar) \log (n \sigma)}$, $\Oh{\sigma K (t_\backward + t_\lastchar) \log (n \sigma) + \sigma^2 t_\backward}$ and $\Oh{t_\lastchar + 1}$ time, respectively, when $v$ is not a node in the original graph.

For \(\sigma = \Oh{1}\), our bounds are summarized in the following theorem.  We will provide more details in the full version of this paper.

\begin{theorem}
\label{thm:bounds}
When \(\sigma = \Oh{1}\), we can store a variable-order de Bruijn graph in $\Oh{n \log K}$ bits on top of the BOSS representation, where $n$ is the number of nodes in the $K$th-order de Bruijn graph, and support \forward\ and \backward\ in $\Oh{\log K}$ time and \lastchar\ in $\Oh{1}$ time.  We can also use $\Oh{n}$ bits on top of the BOSS representation, at the cost of using $\Oh{K \log n / \log \log n}$ time for \forward\ and \backward.
\end{theorem}

\section{Experiments}
\label{sec:experiments}

We have implemented our data structure on top of an efficient implementation of the BOSS single-$K$ data structure\footnote{The implementation is released under GPLv3 license at \url{http://github.com/alexbowe/cosmo}}..
Both structures make use of the SDSL-lite software library\footnote{\url{https://github.com/simongog/sdsl-lite}}.
Our test machine was a server with a 16-core 2.13Ghz Xeon E7-4308 CPU and 128 GB RAM running Fedora Linux 20.
Each experiment was run on a single core, and was repeated three times with the mean values reported.

\paragraph{Test Data.}
Our first data set consists of approximately 27 million paired-end 100 character reads (strings) from {\em E.~coli} (substr.  K-12). It was obtained from the NCBI Short Read Archive (accession ERA000206, EMBL-EBI Sequence Read Archive). The total size of this data set is around 6GB on disk.
The second data set is 36 million 155 character reads from the Human chromosome 14 Illumina reads used
in the GAGE benchmark\footnote{\url{http://gage.cbcb.umd.edu/}}, totalling 6 GB on disk.

\paragraph{Experiments.}


We first used DSK~\cite{dsk} on each data set to find the unique $K$-mers.
It is usual to have DSK ignore low-frequency $K$-mers (as they may result from sequencing errors). 
However, removing such $K$-mers may result in the removal of some shorter $k$-mers that would 
otherwise have an acceptable frequency.  We therefore set the frequency threshold to $1$ (accepting 
all $K$-mers). 
A value of $K = 27$ was chosen
for the {\em E.~coli} data, and $K = 55$
for the Human chromosome $14$ data, as these values produced good assemblies
in previous papers (see, e.g.,~\cite{paul}).
DSK took 28 and 58 minutes to run on the {\em E.~coli} and human data sets, respectively.

The $K$-mers from DSK (and their reverse complements) were then sorted using an in-memory, 
CPU-based radix sort.
The BOSS structure and $L^{*}$ vector were then built using indexes from SDSL-lite. 
Construction times and structure sizes are shown in Table~\ref{tab:nav-time}.
While the variable-$K$ BOSS structure is around $30\%$ slower to build, and $2.6$ to $3.4$ times larger 
than the standard BOSS structure, this is clearly much faster and less space consuming than building 
$K$ separate instances of the BOSS structure.


To measure navigation functions $\forward$ and $\backward$ we took the mean time 
over 20,000 random queries. For the variable-$K$ graph, the $k$ values for each node 
were chosen randomly between $8$ and $K$. 
Results are shown in Table~\ref{tab:nav-time}. 
The new structure makes the $\forward$ operation 2-3 times slower for $k < K$, though we 
note that for $k=K$ $\forward$ time is identical.
The $\backward$ operation is much slower in the new structure, but is much less frequently 
used than $\forward$ in assembly algorithms.
We also measured $\lastchar$, which took only fractions of a nanosecond on both structures.

To see how fast the order can be changed, we timed $\shorter$ and $\longer$ for short (1 symbol) 
and long (4 symbols) changes in the order $k$. Our experiments show that in practice changing 
order by a single symbol ($\shorter_1$ and $\longer_1$) is a cheap operation, taking around the
same time as following an outgoing edge in the original BOSS structure. For longer changes in 
order, the time for $\shorter$ is fairly stable ($\shorter_1$ and $\shorter_4$ take roughly the
same time), whereas $\longer_4$ takes significantly longer than $\longer_1$. This is because
$\longer$ must compute a set of nodes, and the size of that set grows roughly exponentially with
the change in order. 



\begin{table}[ht]
\begin{tabularx}{\textwidth}{@{\extracolsep{\fill} } r  c  c   c  c }
						& \multicolumn{2}{c}{{\em E.~coli}} 		& \multicolumn{2}{c}{Human chromosome 14} \\
						\cline{2-5}
   						& BOSS 		& variable-$K$ BOSS			& BOSS  		&  variable-$K$ BOSS  \\
\hline
Construction (mins) & 19  & 25 (1.32{\sf x}) & 153 & 203 (1.33{\sf x}) \\
Size (MB)  & 163 & 420 (2.58{\sf x}) & 414 & 1416 (3.42{\sf x})\\
\hline
$\forward$ ($\mu$s)  & 28.28 & 80.45  (2.84{\sf x})  & 45.37 & 88.03 (1.94{\sf x})\\
$\backward$ ($\mu$s) & 40.38 & 303.75 (7.52{\sf x}) & 55.19 & 303.21 (5.49{\sf x})\\
$\shorter_1$ ($\mu$s) & N/A & 37.20  & N/A & 46.58 \\
$\shorter_4$ ($\mu$s) & N/A & 45.80  & N/A & 51.59 \\
$\longer_1$ ($\mu$s) & N/A & 83.96  & N/A & 79.90 \\
$\longer_4$ ($\mu$s) & N/A & 1304.54  & N/A & 1837.61 \\
$\maxlen$ ($\mu$s) & N/A & 7  &  N/A & 7 \\
$\maxlen_c$ ($\mu$s) & N/A & 29 & N/A & 33\\
\hline

\end{tabularx}
\caption{Construction time and structure size (top), and mean time taken for each navigation operation (lower), for both data sets and structures. Cells marked ``N/A'' for BOSS indicate operations not possible with that structure.}
\label{tab:nav-time}
\end{table}

\section{Conclusion}\label{sec:conclusion}

We have described a method for efficiently representing multiple de Bruijn graphs 
of different orders in a single succinct data structure. As well as the usual graph 
traversal operations, the data structure supports new operations which allow the order 
of the de Bruijn graph to be changed on the fly. This data structure has the potential 
to greatly improve the memory and space usage of current state-of-the-art assemblers 
that build the de Bruijn graph for multiple values of $K$, and ultimately allow those
assemblers to scale to large, eukaryote genomes. The integration of our new data 
structure into a real assembler is thus our most pressing avenue for future work.

\let\oldbibliography\thebibliography
\renewcommand{\thebibliography}[1]{%
  \oldbibliography{#1}%
  \setlength{\itemsep}{0pt}%
}
\bibliographystyle{abbrv}
\bibliography{dbg}
\end{document}